\documentclass[useAMS,usenatbib]{mnras}
\usepackage[fleqn]{amsmath}
\usepackage{graphicx}	

\newcommand{\der}{{\rm d}}

 \newcommand{\cc}{_{\rm c}}

 \newcommand{\acc}{^{\rm acc}}

 \newcommand{\h}{_{\rm h}}

\newcommand{\ta}{^{\rm nst}} \newcommand{\clR}{R_{\rm s}} 
\newcommand{\clM}{M_{\rm s}} 
\newcommand{\clQ}{Q_{\rm s}}  
 
\newcommand{\clc}{c_{\rm s}}  
 \newcommand{\per}{_{\rm per}}
 \newcommand{\tr}{^{\rm tr}}

 \newcommand{\dc}{_{\rm dDM}}

\newcommand{\maxi}{_{\rm max}} 
\newcommand{\modot}{M$_\odot$\ } \newcommand{\modotc}{M$_\odot$}
 
\newcommand{\ii}{_i}

\newcommand{\E}{_{\rm E}} \newcommand{\LL}{_{\rm L}}
 
\newcommand{\beq}{\begin{equation}} \newcommand{\f}{^f} \newcommand{\eeq}{\end{equation}}
 \newcommand{\beqa}{\begin{eqnarray}}
\newcommand{\eeqa}{\end{eqnarray}} \newcommand{\lav}{\langle}
\newcommand{\rav}{\rangle} 
\newcommand{\vir}{_{\rm vir}} 
 
 \newcommand{\ff}{^{\rm f}} \newcommand{\PA}{^{\rm PA}}

\newcommand{\fin}{^{\rm stp}}  \newcommand{\rfin}{^{\rm fin}} 
 
   \newcommand{\uc}{^{\rm c}}
 \newcommand{\DF}{_{\rm DF}}

\begin{document}

\title[Dynamical Friction]{An Accurate Comprehensive Approach to Substructure: \\IV. Dynamical Friction}


\author[Salvador-Sol\'e, Manrique \& Rocamora]{Eduard
  Salvador-Sol\'e$^1$\thanks{E-mail: e.salvador@ub.edu}, Alberto Manrique and Andreu Rocamora
  \\Dept. F\'{\i}sica Qu\`antica i Astrof\'\i sica, Institut de Ci\`encies del Cosmos (ICCUB), Facultat de F\'\i sica, Universitat de Barcelona,\\Mart\'\i \ Franqu\`es, 1, E08028 Barcelona, Spain}


\maketitle
\begin{abstract}
In three previous Papers we analysed the origin of the properties of halo substructure found in simulations. This was achieved by deriving them analytically in the peak model of structure formation, using the statistics of nested peaks (with no free parameter) plus a realistic model of subhalo stripping and shock-heating (with only one parameter). However, to simplify the treatment we neglected dynamical friction (DF). Here, we revisit that work by accounting for DF. That is also done in a fully analytic manner that avoids the numerical integration of the subhalo orbital motion. We obtain very simple expressions for the abundance and radial distribution of subhaloes of different masses that disentangle the effects of DF from those of tidal stripping and shock-heating. These analytic expressions reproduce and explain the results of simulations and allow one to extend them to haloes of any mass, redshift and formation times in any desired cosmology. 
\end{abstract}

\begin{keywords}
methods: analytic --- gravitation --- cosmology: theory, dark matter --- galaxies: haloes, structure 
\end{keywords}


\section{INTRODUCTION}\label{intro}

Dynamical friction (DF) plays an important role in the evolution of many self-gravitating systems such as massive stars in young galaxies (\citealt{Rea21}), spiral arms in disc galaxies (\citealt{Sell21,C23}), stars in globular clusters (\citealt{SGH21,Bea23}), globular clusters in dwarf galaxies (\citealt{Li21,Bo21,Shea2}), stars around supermassive black holes \citep{Gea23}, black hole binaries (\citealt{Beea23}), super massive black holes in galaxies (\citealt{Ma21,Rea23,DGea23}) and primordial black holes (\citealt{Su21,SB23}). Unfortunately, the absence of a fully analytic treatment of DF complicates their modelling. 

That is the case in particular of substructure in cold dark matter (CDM) haloes (e.g., \citealt{LC93,TB01,Bea02}). Due to he complexity of the problem, involving the subhalo aggregation history and their evolution through tidal stripping, shock heating and DF as they orbit inside the host haloes, the usual way to address it has been by means of high-resolution cosmological $N$-body simulations (e.g. \citealt{Dea07,Sea08a,Aea09,Eea09,BK10,Gi10,Kea11,Gea11,Gea12,Oea12,Lea14,Cea14b,Iea20}), recently taking into account the hydrodynamics of gas (e.g. \citealt{Rich20,Fea20,Fea20b,Hell16,Bea16,Bea20}). But this approach is very CPU time-consuming, so the properties of substructure are only known for a few haloes with specific aggregation histories. In addition, simulations do not facilitate a detailed understanding of how these properties are set. This is why (semi) analytic models (e.g. \citealt{TB01,Fea02,ZB03,S03,L04,OL04,TB04,Pea04,vdB05,Zeabis05,KB07,Gi08,Aea09,Bea13,PB14,Ji16,Gfea16,vdB16,vdB18,GB19,Jea21}) have also been used. Unfortunately, in the absence of an analytic treatment of DF (see the work in this direction by \citealt{BD23}), analytic models must integrate the subhalo orbital motion, which is similarly CPU time-consuming. What is worse, the numerical integration of orbits deprives the models from their main reason to be: finding simple analytic expressions facilitating the comprehension of the problem and describing the general case. 

In \citep{I,II,III}, hereafter Papers I, II and III, respectively, we built a very detailed analytic model of halo substructure in the peak model, based on the powerful {\it ConflUent System of Peak trajectories} (CUSP) formalism \citep{Mea95,Mea96,Mea98} having also allowed us to derive analytically all the remaining inner halo properties (\citealt{SM19} and references therein), their clustering \citep{SM24,Sea24} and angular momentum growth \citep{SM25}). In those Papers we were able to reproduce and explain the abundance or mass function (MF) and radial distribution of accreted non-evolved subhaloes (Paper I) and evolved ones through the action of tidal stripping and shock-heating (Paper II) in both purely accreting and ordinary haloes of different masses, redshifts and aggregation histories (Paper III). However, to facilitate the analytic treatment we neglected DF, so the results obtained only held for subhaloes less massive than $\sim 10^{-4}$ the host mass. 

In this Paper we remedy that limitation. We revisit the study by including DF, treated in a fully analytic manner. This allows us to obtain simple analytic expressions for the subhalo MF and radial distribution showing how DF alters the results derived in Papers II and III. This way, we disentangle the role of all the different processes shaping the properties of halo substructure found in simulations and extend them to haloes of any mass, redshift, and formation time in any desired CDM cosmology. 

The layout of the Paper is as follows. In Section \ref{DynFric} we study the effect of DF on individual subhaloes. In Section \ref{population} we implement the results to the entire subhalo population of purely accreting haloes and real ones having undegone major mergers. The summary and concluding remarks are given in Section \ref{sum}. 

To facilitate the comparison of the results obtained here with those inferred without DF in previous Papers (and found in simulations by \citealt{Hea16}), all Figures shown throughout the Paper assume, unless otherwise stated, Milky Way (MW) haloes, i.e. with virial mass $M\h=2.2\times 10^{12}$ \modotc,\footnote{The virial mass of a halo is defined as usual: the mass inside the radius encompassing an inner mean density equal to $\Delta\vir(z)$ \citep{BN98} the mean cosmic density.} endowed with the NFW density profile \citep{NFW95} with concentration $c=12$ in the {\it WMAP7} cosmology \citep{Km11}, with $(\Omega_\Lambda,\Omega_{\rm m},h,n_{\rm s},\sigma_8,\Omega_b)=(0.73,0.27,0.70,0.95,0.81,0.045)$.

\section{DF on Individual Subhaloes}\label{DynFric}

There are in the literature two different mechanisms referred to DF. One is that caused by the {\it local wake} produced by light particles of a continuous medium scattered behind an object moving inside it. This mechanism, introduced by \citet{Cha43}, causes the velocity loss and orbital decay of subhaloes \citep{Be89,Mu93,CP98,Cea99,BT08}. But the torque produced by the long-scale resonant interaction of a moving subhalo with the host halo also contributes to its orbital decay \citep{W83,TW84,W86,W89,Cea09,OB16,GCea19,CGC20,Tea21}. This is why this latter mechanism is also called {\it global mode} DF even though it does not really behave as a friction. \citet{Tea21} showed that, when there is one only very massive subhalo, this global mode DF is stronger than the former local one. However, very massive ones undergo very strong DF of the former kind and fall into the centre of the host halo in one (long) orbit and disappear, while the global mode DF does not take place during the first orbit of a subhalo \citep{Tea21}. Thus, we concentrate from now on in the effects of local DF. 

The equation of motion of a subhalo with mass $\clM$ orbiting within a spherical halo of mass $M\h$ at some cosmic time $t\h$, subject to the action of the local wake DF is
\beq
\ddot{\bf r} = -\nabla \Phi (r) - A(v,r,\clM) \dot {\bf r}, 
\label{motion}
\eeq
where ${\bf r}$ is the position vector of the subhalo with origin at the centre of the halo, $\Phi(r)$ is the gravitational potential and $A$ is the (positive) DF coefficient \citep{Cha43}, well approximated in finite spherical systems by (e.g. \citealt{Pea10,Jea21}) 
\beq
A(v,r,\clM)=4\pi\, G^2\, \clM\,\rho(r)\, f\dc(r)\, \ln \Lambda\, \frac{F(<v)}{v^3}.
\label{A1}
\eeq
For simplicity in the notation, here and in what follows, we skip the arguments $M\h$ and $t\h$ referring to the mass and cosmic time of the host halo, in all quantities that depend on them. They will only be included in Section \ref{ordinary} when dealing with haloes of different masses and times. 

In equation (\ref{A1}), $G$ is the gravitational constant, $\ln\Lambda$ is the so-called Coulomb logarithm (see the discussion below), $v$ is the modulus of $\dot{\bf {r}}$, $F(<v)$ is the fraction of particles with relative Maxwellian-distributed velocities less than $v$, equal to erf$(X)-(2/\sqrt{\pi})X\exp(-X^2)$, where $X\equiv\sqrt{3/2}\,v/\sigma(r)$, $\sigma (r)$ and $\rho(r)$ are the isotropic (3D) velocity dispersion and density profiles of the host halo and $f\dc(r)$ is the diffuse dark matter (dDM) mass fraction (i.e. the fraction of dark matter outside subhaloes). As mentioned, as long as haloes are accreting, they grow inside-out \citep{Sea12a}, so the local density and velocity dispersion at any fixed radius stay essentially unaltered. Strictly speaking, the dDM fraction $f\dc(r)$ slightly increases with time as subhalo stripping progresses \citep{II} and the density profile of haloes slightly deepens as a consequence of DF acting on massive subhaloes. However, for simplicity, these minor effects are ignored. 

The Coulomb logarithm is a fudge factor introduced to adapt the \citet{Cha43} formula originally derived by \citet{Cha43} for homogeneous infinite systems to finite ones with some particular geometry. There are different more or less complicate forms in the literature for that factor (e.g. \citealt{Rea06,G10,TB01,PB05,AB07,Pea10}). From now on we will assume for simplicity that it does not depend on $r$ and $v$, with a constant value of 2.1 as shown to reasonably reproduce the results of simulations for spherical pure CDM haloes endowed with density profiles of the typical NFW form \citep{PB05,AB07,Pea10}. Note also that in equation (\ref{A1}) we have neglected the DF caused by less massive subhaloes as it is much less marked than that caused by dDM. 

\begin{figure*}
\centerline{\includegraphics[scale=1.20,bb= 45 0 290 200]{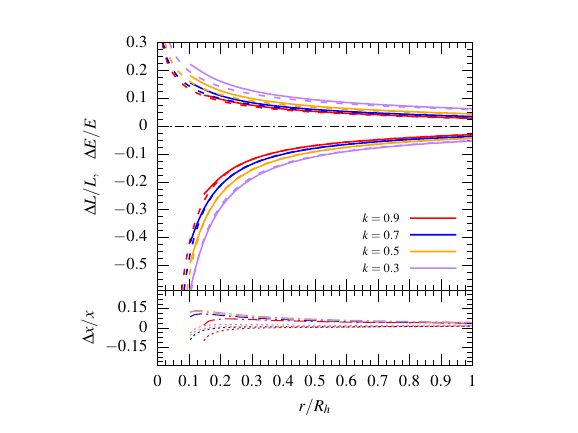}
\includegraphics[scale=1.20,bb= 80 0 240 200]{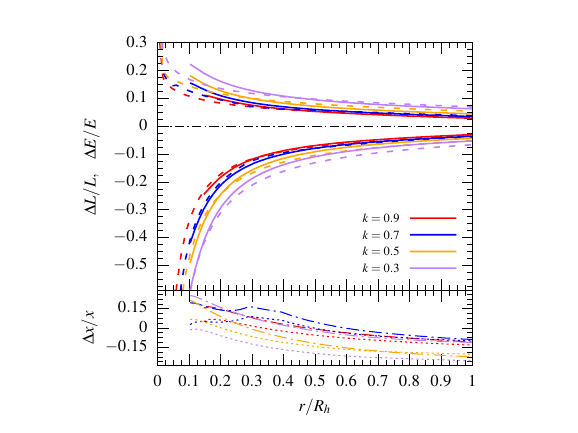}}
\caption{Approximate relative orbital energy $E$ and angular momentum $L$ increments (positive and negative, respectively) of subhaloes of mass $\clM=10^{-2}M\h$ as a function of their initial apocentric radius $r$ (scaled to the virial radius $R\h$ of the halo) and the parameter $k$ measuring the square of their initial tangential velocity scaled to the circular velocity (dashed lines) in current haloes with MW-mass $M\h$, compared to the exact results obtained by integration over real orbital motions with DF (solid lines). The black dot-dashed line marks the zero baseline. In the bottom panel we plot the relative differences between the approximations (dotted lines for $\Delta E/E$ and dot-dashed lines for $\Delta L/L$) and the exact solutions (the zero baseline). Left panel: Approximate $\Delta E/E$ and $\Delta L/L$ values obtained by integration over virtual orbits with no DF. Right panel: Approximate purely analytic $\Delta E/E$ and $\Delta L/L$ values (i.e. obtained with no integration).\\
(A colour version of this Figure is available in the online journal.)}
\label{f0}
\end{figure*}

\subsection{One Orbit}\label{moderate}

Next, in a first step we calculate the effect of DF alone and, in a second step, its combined effect with tidal stripping and shock-heating. 

\subsubsection{DF only}

Multiplying equation (\ref{motion}) by $\dot {\bf r}$ and integrating over one orbit, we obtain
\beq
E(r\ff)=E(r)+\Delta E,
\eeq
where $E$ is the total orbital energy of the subhalo, $r$ and $r\ff$ are the initial and final apocentric radii, respectively, and
\beq
\Delta E=\! -\clM \!\int_0^T \!\!A[v(t),r(t),\clM] v^2(t) \der t\equiv\! -\clM A\E\! \int_0^T \!\! v^2 \der t,
\label{Einvariant}
\eeq
with $T$ being the orbital period, dependent on the subhalo apocentric radius, (tangential) velocity at that radius and mass, and $A\E$ being the energy-averaged DF coefficient over one orbit. 

On the other hand, since the momentum of a central force is null, the direction of the angular momentum of the subhalo relative to the centre of the halo, ${\bf L}=\clM {\bf r} \times {\bf \dot r}$, with $\ddot {\bf r}$ given by equation (\ref{motion}), is kept constant and the time-derivative of its modulus is simply $- A L$, so, integrating it over one orbit, we find 
\beq
L(r\ff)=L(r)+\Delta L, 
\label{am}
\eeq
where
\beqa
\Delta L = -L \left(1-\exp{\left\{-\int_0^T A
[v(t),r(t),\clM] \der t\right\}}\right)~~~~~~~~\nonumber\\
= -L\int_0^T A[v(t),r(t),\clM] \der t\equiv -LA\LL T,~~~~~~~~~~~~~~~
\label{Linvariant}
\eeqa
with $A\LL$ being the angular momentum-averaged DF coefficient over one orbit. Equation (\ref{Linvariant}) holds to first order in the effects of DF as most equations throughout this Paper, but, for simplicity, we write from now on the symbol $=$ and reserve the symbol $\approx$ for the case of some additional approximation. 

\begin{figure*}\centerline{\includegraphics[scale=1.30,bb= 45 0 240 200]{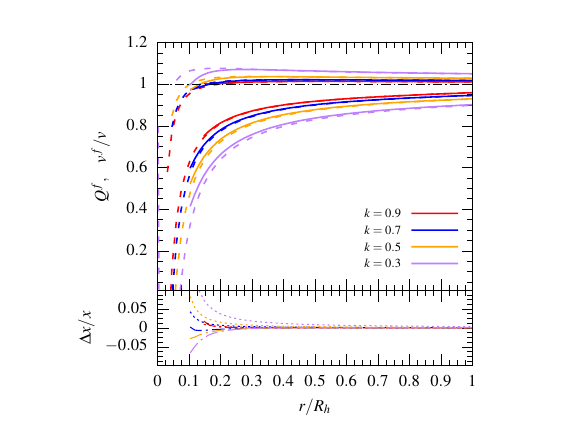}
\includegraphics[scale=1.30,bb= 50 0 230 200]{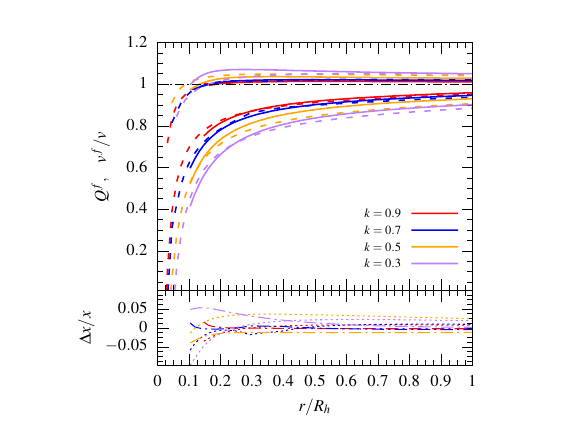}}
\caption{Ratios of final to initial radii (curves below unity) and tangential velocities (curves above unity) at apocentre in one orbit found by numerical integration of the orbital motion of subhaloes (solid lines) and obtained to first order in the relative energy and angular momentum increments, $\Delta E/E$ $\Delta L/L$ (dashed lines), as a function of $r$ for several $k$ values and the same $\clM$, $M\h$ and $t\h$ as in Figure \ref{f0}. Again, the black dot-dashed line marks the zero value. In the bottom panel we plot the relative differences between the approximations (dotted lines for apocentric radii and dot-dashed lines for tangential velocities) and the exact solutions (the zero baseline). Left panels: Results obtained using the leading-order-approximate $\Delta E/E$ and $\Delta L/L$ increments found by integration over virtual orbits with no DF. Right panels: Results obtained using the approximate purely analytic $\Delta E/E$ and $\Delta L/L$ values.\\
(A colour version of this Figure is available in the online journal.)}
\label{f1}
\end{figure*}

In Appendix \ref{App1} we calculate the relative increments $\Delta E/E$ and $\Delta L/L$, positive and negative, respectively, as functions of the initial apocentric radius $r$ and (tangential) velocity $v$ or, equivalently, as functions of $r$ and $k\equiv v^2/[GM(r)/r]$, where $GM(r)/r$ is the squared circular velocity at $r$ (hence, $k\le 1$, with $k=1$ standing for circular orbits). In terms of $r$ and $k$, the initial orbital energy and angular momentum take the form $E=\clM[kGM(r)/(2r)+\Phi(r)]$, where $\Phi(r)$ is the gravitational potential of the halo, and $L=\clM r [k GM(r)/r]^{1/2}$. The calculation is achieved in two different ways: 1) a first one accurate to leading order in the effects of DF, in which the integrals in equations (\ref{Einvariant}) and (\ref{Linvariant}) are carried out over the well-known orbits {\it without DF}, i.e. with no need to solve the equation of motion of subhaloes with DF, and 2) a second less accurate though fully analytic one involving no integral at all. 

Both versions of $\Delta E/E$ and $\Delta L/L$ are compared in Figure \ref{f0} to the exact values of these quantities found by numerical integration over the real subhalo orbit with DF. The solid lines giving the exact $\Delta E/E$ and $\Delta L/L$ values are truncated at $r\la 0.1$. At smaller $r$, subhaloes spiral inwards without reaching any new apocentre (and pericentre), so the definition of the orbital period $T$ as the time between two consecutive apocentres becomes meaningless. (The same situation is found for $k$ close to unity, i.e. quasi-circular orbits; in Figure \ref{f0} that happens at $k>0.9$.) Defining $T$ at those radii as the time spent until the subhalo reaches the halo centre is not a solution because this would yield a large discontinuity in those lines. The best solution is to define the orbital period at those radii ($k$ values) by continuity with the values found at larger $r$ for the same $k$ (at smaller $k$ for the same $r$). This is what we have done for approximate $\Delta E/E$ and $|\Delta L/L|$ values (dashed lines). Had we adopted the same definition for the exact $\Delta E/E$ and $|\Delta L/L|$ values, the comparison of the dashed lines to the solid ones at small $r$ (large $k$) would be similarly good. In what follows we use such an extended orbital period.

The changes in one orbit of the apocentric radius $r$ with (tangential) velocity $v$ to the final ones, $r\ff$ and $v\ff$, can be calculated to first order in $\Delta E/E$ and $\Delta L/L$, writing the particle velocity at apocentre in terms of the radius and the angular momentum and Taylor expanding to first order the potential at the final radius $\Phi(r\ff)$ around the initial one $r$. This leads to a cubic equation for the ratio $Q\ff\equiv r\ff/r$, whose result is
\beq
Q\ff\!=\!1+\frac{k}{1-k}\!\left[\frac{S(k,r)}{2}\frac{\Delta E}{E}(k,r,\clM)-\frac{\Delta L}{L}(k,r,\clM)\right]\!,\label{Qf}
\eeq
with $S(k,r)\equiv 1+2r\Phi(r)/[kGM(r)]=1-2/k \ln [1+c(r)]/f[c(r)]<0$, leading to 
\beqa
\frac{r\ff}{r}=Q\ff(k,r,\clM)\,~~~~~~~~~~~~~~~~~~~~~~~~~~~~~~~~~~~~~~~~~~~~~~~~~~\label{rfr}\\
\frac{v\ff}{v}=\frac{1+\frac{\Delta L}{L}(k,r,\clM)}{Q\ff(k,r,\clM)}.\!~~~~~~~~~~~~~~~~~~~~~~~~~~~~~~~~~~~~~~~~~~~\label{vfv}
\eeqa

In Figure \ref{f1} we compare the ratios $r\ff/r$ and $v\ff/v$ obtained from the two approximate versions of $\Delta E/E$ and $\Delta L/L$ to the exact values obtained by numerical integration. As can be seen, the most accurate estimates of $\Delta E/E$ and $\Delta L/L$ lead to $r\ff/r$ and $v\ff/v$ ratios that almost fully recover the exact ones at all $r$ and $k$. But, even the less accurate fully analytic estimates of $\Delta E/E$ and $\Delta L/L$ give very good results, so we adopt them in what follows.

Certainly, equations (\ref{rfr}) and (\ref{vfv}) hold to first order in the quantities $\Delta E/E$ and $\Delta L/L$, which are proportional to $\clM$ (see App.~\ref{App1}), so those values of $r\ff/r$ and $v\ff/v$ might not be accurate enough for massive subhaloes suffering strong DF. But this is not the case. As long as $r\ff/r$ is non-null, equations (\ref{rfr}) and (\ref{vfv}) give fairly good estimates of the exact $r\ff/r$ and $v\ff/v$ values regardless of the subhalo mass (see Figs.~\ref{f0} and \ref{f1}). While, when $r\ff/r$ vanishes, the subhalo falls into the halo centre, merges with the central dark matter lump and disappears (see Paper II), so we must no longer monitor its orbital motion. Therefore, equations (\ref{rfr}) and (\ref{vfv}) can be safely applied for all subhalo masses.

\subsubsection{DF Combined with Stripping and Shock-Heating}\label{stripping}

Although tides operate over the entire subhalo orbit, their most marked effect takes place at the pericentric radius, $r\per$. Before that, the softly stripped material stays close to the subhalo, so the DF produced on the subhalo and its stripped matter is the same as if the subhalo had not been stripped. In contrast, at pericentre the (cumulative and new) stripped material substantially separates from the subhalo due to the shock heating produced at that point, so, its way back to the apocentre, DF acts on the subhalo according to its new mass. Until reaching the new apocentre, the subhalo is kept essentially unchanged because the stripped subhalo has no time to relax and with that shape the stripping is not so marked as at pericentre. Only when the subhalo is near the apocentre has the subhalo enough time to relax and its structure acquires a new equilibrium configuration with which it begins a new orbit. 

As shown in Paper II, for subhaloes at the apocentric radius $r$, with velocity $v$, mass $\clM$ and a NFW density profile with concentration $\clc$, the ratio $\clQ(v,r,\clM)\equiv \clR\tr(v,r,\clM)/\clR$ between the final truncated radius (after having been stripped and shock-heated at pericentre) and the initial radius satisfies
\beq
\frac{f(\clc \clQ)}{f(\clc)(\clQ)^3}= 
\frac{f\left[c(r)Q\per(v,r,\clM)\right]}{f[c(r)]Q\per^3(v,r,\clM)},
\label{mtr2}
\eeq
with $f(x)=\ln(1+x)-x/(1+x)$. The truncated mass $\clM\tr$ satisfies in turn
\beq
\frac{\clM\tr}{\clM}=\frac{f\left(\clc \clQ\right)}{f(\clc)}=\frac{f\left[c(r)Q\per(v,r,\clM)\right]}{f[c(r)]Q\per^3(v,r,\clM)}(\clQ)^3,
\label{second}
\eeq 
where $c(r)$ is the concentration of that inner part of the halo with mass $M(r)$, i.e. its total radius $r$ over the constant scale radius $r_0$ (accreting haloes grow inside-out), hence, $c(r)=rc\h/R\h<c\h$, with $c\h=R\h/r_0$ being the concentration of the entire halo, and $Q\per$ is defined as $r\per/r$. In the absence of DF, $r\per$ and $Q\per$ do not depend on $\clM$, so, since $\clc$ is weakly dependent on $\clM$, $\clQ$ is also essentially independent of $\clM$ (see eq.~[\ref{mtr2}]). However, in the presence of DF, the pericentric radius and the ratio of pericentric-to-apocentric radii, hereafter denoted as $r\per\ff$ and $Q\per\ff$, respectively, change depending on $\clM$ and, consequently, the ratio of initial-to-final subhalo radii, hereafter denoted as $\clQ\ff$, does too. That is the only difference introduced by DF in the stripping model of Paper II. 

To calculate the mass $M\ff$ of the stripped subhalo in the presence of DF we need the ratio $Q\per\ff=r\per\ff/r$. The relation between the modified and original pericentric radii, $r\per\ff$ and $r\per$, is obviously the same as between the modified and original apocentric radii (eq.~[\ref{rfr}]) after one orbit. But the deviation of $r\per\ff$ from $r\per$ {\it after half one orbit} (i.e. since the previous apocentre at $r$) is given by $r\per\ff/r\per$ equal to $Q\ff$ (eq.~[\ref{Qf}]) for $\Delta E/E$ and $\Delta L/L$ corresponding to half the period $T$ (or, equivalently, corresponding to the whole period though half the mass $\clM$; see App.~\ref{App1}). And, using the approximation $r\per/r\approx \tilde k\equiv k(1+\sqrt{1+8/k})/4$ (see App.~\ref{App1}), we obtain
\beq
Q\per\ff(k,r,\clM)\equiv\frac{r\per\ff}{r}
\approx \tilde k\,Q\ff(k,r,\clM/2).
\label{Qper}
\eeq
To write equation (\ref{Qper}) we have taken into account that $Q\ff$ (eq.~[\ref{Qf}]) for $\Delta L/L$ and $\Delta E/E$ corresponding to half the period $T$ and mass $\clM$ equals $Q\ff$ for the whole period $T$ and the mass $\clM/2$ (see eqs.~[\ref{AE2}] and [\ref{AL2}]). Note that the effective values of $A\E$ and $A\LL$ averaged over half the period (from apocentre to pericentre) are indeed the same as averaged over the full orbit (see eqs.~[\ref{Einvariant}] and [\ref{Linvariant}]). 

After stripping at pericentre, the subhalo ends its orbit with the stripped mass $\clM\ff$. Since half the orbit is carried with the original mass $\clM$ and the other half with the stripped mass $\clM\ff$, the changes produced in the apocentre after completing the orbit with stripping at the pericentre coincide with the arithmetic mean of those produced with DF only over the orbit of the subhalo with the two masses. Lastly, when the subhalo reaches the apocentre, it settles in a new equilibrium configuration, with NFW density profile though a somewhat larger concentration $\clc\ff$ so that, in its next orbit, it will be further stripped and heated (see Sec.~\ref{multiple}). In the impulsive approximation and no DF, the new concentration is related to the original one through (Paper II)
\beq
\frac{h(\clc\tr)}{h(\clc)}=\kappa \left(\frac{\clM\tr}{\clM}\right)^{\beta+5/6},
\label{rat2} 
\eeq
where $h(c)\equiv f(c)(1+c)/\{c^{3/2}[3/2-s^2(c)]^{1/2}\}$, being $s^2(c)$ the isotropic 3D velocity variance $\sigma^2$ scaled to $cf(c)GM/R$ of a halo with virial mass $M\h$, virial radius $R\h$ and concentration $c$. And, in the presence of DF, the same derivation leads to equation (\ref{rat2}) with $\clc\tr$ and $\clM\tr$ replaced by $\clc\ff$ and $\clM\ff$ resulting from the combined action of DF and stripping plus shock-heating. Constants $\kappa=0.77$ and $\beta=-1/2$ give very good fits to specific numerical experiments (Paper II).

\subsection{Multiple Concatenated Orbits}\label{multiple}

To find the mass and radius of subhaloes at the final time $t\h$ we must calculate in an iterative way their changes produced in successive orbits since the time $t$ they are accreted. To do this we need first to calculate, following Paper II, the values at $t$ of all compelling quantities mentioned in Section \ref{stripping}.

The inside-out growth of accreting haloes \citep{Sea12a} guarantees that the mass $M(t)$ of the progenitor halo at $t$ coincides with the mass $M(r)$ in the final halo inside the radius $r$ reached by the progenitor at that time. Thus, the equality $M(t)=M(r)$, where $M(t)$ is the typical mass at $t$ of accreting haloes with $M\h$ at $t\h$ provided by CUSP \citep{SM19} and $M(r)$ is the NFW mass profile of such haloes (reproduced by CUSP; \citealt{Sea23}), is an implicit equation for $r(t)$. 

Interestingly, $r$ is precisely the apocentric radius of subhaloes accreted at $t$ (Paper I). Indeed, after reaching turnaround, subhaloes fall onto the progenitor halo and bounce, giving rise to a relaxation period during which they cross back and forth the central progenitor halo and next falling shells. This ordered crossing causes them to lose part of the orbital energy (see \citealt{Sea12a} for details), so their orbits gradually shrink. As their phases with respect to that of outer shells become increasingly uncorrelated, the energy transfer brakes until the orbits stop shrinking and stabilise. Thus, the apocentric radius of those newly virialised subhaloes marks the new instantaneous virial radius $r$ of the progenitor halo, while their tangential velocity $v$ at $r$ is randomly distributed, independently of their mass \citep{Jea15}, according to the distribution function given in Paper II. 

In these circumstances all quantities referring to subhaloes accreted at $t$ can be derived using the equations given in Section \ref{stripping} with `initial' values (previous to the subhalo stripping and shock-heating suffered during the virialisation process) denoted with index nst and `final' values at $t$ denoted with no index. Equations (\ref{mtr2}) and (\ref{second}), with final mass $\clM$, initial concentration $\clc\ta$ given by the $M$-$c$ relation corresponding to $\clM\ta$ at $t$, the concentration $c(r)=r/r_0$ of the progenitor halo and the ratio $Q\ff\per$ equal to twice the velocity-averaged ratio of subhaloes at $t$,\footnote{The virialisation of spherical homogeneous protohaloes \citep{BN98} and non-homogeneous triaxial ones as well \citep{SM19} causes the system to contract a factor 2 since turnaround. Thus, that is the contraction of the subhalo apocentric radius. However, their pericentric radius stays essentially unaltered because the density profile of accreting haloes does not essentially change during that time. Consequently, $r\per/r$ increases approximately a factor two.} are two implicit equations for the initial mass $\clM\ta$ and final scaled truncation radius $\clQ$. Then, plugging the values of $\clM/\clM\ta$ and $\clc\ta$ in equation (\ref{rat2}), we obtain the final concentration $\clc$ of subhaloes at $t$. As we will see in Section \ref{population}, this approximate procedure is enough to obtain the right radial abundance of the evolved subhaloes.

To carry out the iterative derivation below it is convenient to denote the starting quantities $r$, $v$, $k$, $\clM$, $\clQ$ and $\clc$ of subhaloes accreted at $t$ with index zero. 

After one orbit, these subhaloes reach a new apocentric radius $r_1$ and velocity $v_1$ given by equations (\ref{rfr})-(\ref{vfv}) for the mass $(M_0+M_1)/2$, with the mass $M_1$ given by equation (\ref{second}), factor $Q_1$ given by equation (\ref{mtr2}) and concentration $c_1$ given by equation (\ref{rat2}). Then, they start a new orbit and so on.

In general, after the $i+1$ orbit, subhaloes with initial apocentric radius $r\ii$ and tangential velocity $v\ii$ end up with the values $r_{i+1}$ and $v_{i+1}$ given by 
\beqa
\frac{r_{i+1}}{r\ii}=\frac{1}{2}\,\left[Q\ff(k\ii,r\ii,M\ii)+Q\ff(k\ii,r\ii,M_{i+1})\right]~~~~~~~~~~~~~~\label{ri}\\
\frac{v_{i+1}}{v\ii}\!=\!\frac{1}{2}\!\left[\!
\frac{1+\frac{\Delta L}{L}(k\ii,r\ii,M\ii)}{Q\ff(k\ii,r\ii,M\ii)}+
\frac{1+\frac{\Delta L}{L}(k\ii,r\ii,M_{i+1})}{Q\ff(k\ii,r\ii,M_{i+1}}\right],~\label{vi}
\eeqa
where 
\beq 
\frac{M_{i+1}}{M\ii}=\frac{f\left(c\ii Q_{i+1}\right)}{f(c\ii)},
\label{Mi}
\eeq
with $Q_{i+1}$ and $c_{i}$ given by
\beqa
\frac{f(c\ii Q_{i+1})}{f(c\ii) Q^3_{i+1}}= 
\frac{f\!\left[c(r\ii)\tilde k\ii  Q\ff(k\ii,r\ii,M\ii/2)\right]}{f[c(r\ii)]\tilde k\ii^3 [Q\ff(k\ii,r\ii,M\ii/2)]^3}\,~~~~~~~~~~~~~~~~~~
\label{mtr2bis}\\
\frac{h(c\ii)}{h(c_{i-1})}=\kappa\left(\frac{M_{i}}{M_{i-1}}\right)^{\beta+5/6}.~~~~~~~~~~~~~~~~~~~~~~~~~~~~~~~~~~~~
\label{ratot3} 
\eeqa
Strictly speaking, when $M\ii$ is less than the inner mass of the host halo at the pericentre, the subhalo no longer suffers stripping and shock-heating (see Paper II). But, for simplicity, we will not make this distinction here, though the results in Section \ref{ordinary} are derived from those without DF obtained in Paper II accounting for it. 

The previous recursive procedure leads to the final apocentric radius $r\ff$, tangential velocity $v\ff$ and subhalo mass $\clM\ff$,  related to their respective initial values at accretion through 
\beqa
\frac{r\ff(k,r,\clM)}{r}=\prod_{i=0}^{\nu}\frac{r_{i+1}}{r\ii}~~~~~~~~~~~~~~~~~~~~~~~~~~~~~~~~~~~~~~~~~~\label{lastr}\\
\frac{v\ff(k,r,\clM)}{v}=\prod_{i=0}^{\nu}\frac{v_{i+1}}{v\ii},~~~~~~~~~~~~~~~~~~~~~~~~~~~~~~~~~~~~~~~~~\label{lastv}\\
\frac{\clM\ff(k,r,\clM)}{\clM}=\prod_{i=0}^{\nu}\,\frac{M_{i+1}}{M\ii}~~~~~~~~~~~~\,~~~~~~~~~~~~~~~~~~~~~~~~~~\label{lastM}
\eeqa
where $\nu$ is the maximum integer $i$ satisfying the condition $\sum_0^\nu T(k\ii,r\ii,M\ii)< t\h -t(r)$ and (see eq.~[\ref{Qf}])
\beqa
\frac{r_{i+1}}{r\ii}=1+\frac{\Delta r\ii}{r\ii}~~~~~~~~~~~~~~~~~~~~~~~~~~~~~~~~~~~~~~~~~~~~~~~~~~~~\label{rilin}\\
\frac{v_{i+1}}{v\ii}=1-\frac{\Delta r\ii}{r\ii}+\frac{\Delta L}{L}(k\ii,r\ii,\widetilde M\ii)~~~~~~~~~~~~~~~~~~~~~~~~~~~~~\label{vilin}\\
\frac{M_{\rm i+1}}{M\ii}=\frac{M'_{i+1}}{M\ii}\left(1+\frac{\Delta M\ii}{M\ii}\right),~~~~~~~~~~~~~~~~~~~~~~~~~~~~~~~~~~~\label{Milin}
\eeqa
with
\beqa
\frac{\Delta r\ii}{r\ii}\!\equiv\! \frac{k\ii}{1-k\ii}\!\!\left[\frac{S(k\ii,r\ii)}{2}\frac{\Delta E}{E}(k\ii,r\ii,\widetilde M\ii)\!-\frac{\Delta L}{L}(k\ii,r\ii,\widetilde M\ii)\!\right]\label{rilin2}\\
\frac{\Delta M\ii}{M\ii}= J(k,r)\frac{\Delta r\ii}{r\ii}\qquad~~~~~ J(k,r)\equiv \frac{1-g[c(r)\tilde k]}{2[f(\clc \clQ)-1/3]}.~
\label{Milin2}
\eeqa
In equations (\ref{vilin}) and (\ref{rilin2}), $\widetilde M\ii\equiv (M\ii+M_{\rm i+1})/2$ can be replaced, at the same order of approximation, by $M\ii$. Note tha, contrarily to $r_{i+1}/r\ii$ and $v_{i+1}/v\ii$, $M_{i+1}/M\ii$ is not written as a small deviation from unity, but from the value (Paper II)
\beq
\frac{M_{i+1}'}{M\ii}=\frac{f(c\ii Q_{i+1}')}{f(c\ii)}
\eeq
independent of $\clM$ that results from tidal stripping and shock-heating only, with $Q_{i+1}'$ being the solution of the implicit equation
\beq
\frac{f(c\ii Q_{i+1}')}{f(c\ii)(Q_{i+1}')^3}=\frac{f[c(r\ii)\tilde k\ii ]}{f[c(r\ii)]\tilde k\ii^3 }.
\label{mt2bis}
\eeq
The reason for this is that the latter may already notably deviate from unity. After some algebra keeping to first order as usual, this leads to equation (\ref{Milin2}), where we have defined the function $g(x)\equiv [x/(1+x)]^2/[3f(x)]$, taken into account that $c\ii\gg 1$ and $Q_{i+1}\sim 1$, so $g(c\ii Q_{i+1})$ is approximately equal to $1/[3f(c\ii Q_{i+1}]$, and used that $f(c\ii Q_{i+1})$ and $g[c(r\ii)\tilde k\ii]$ can be replaced, at the same order of approximation, by $f(\clc \clQ)$ and $g[c(r)\tilde k]$, respectively. 

Finally, in Appendix \ref{App3} we show that equations (\ref{rilin})-(\ref{Milin}) lead to the following relations between the initial and final subhalo radius and mass 
\beqa
\frac{r\ff(k,r,\clM)}{r}=1+\frac{\Delta r}{r}~~~~~~~~~~~~~~~~~~~~~~~~~~~~~~~~~~~~~~~~~~~~\label{radii}\\
\frac{\clM\ff(k,r,\clM)}{\clM}=\frac{\clM\tr(k,r)}{\clM}\left[1+\frac{\Delta \clM}{\clM}\right],~~~~~~~~~~~~~~~~~~~~~~\label{masses}
\eeqa
where
\beqa
\frac{\Delta r(k,r,\clM)}{r}\equiv\frac{r\ff-r}{r}=\sum_{i=0}^\nu \frac{\Delta r\ii}{r\ii}\approx -Y(k)\tilde I_0(r)\clM~~~~~\\
\frac{\Delta \clM(k,r,\clM)}{\clM}\equiv \frac{\clM\ff-\clM}{\clM}=\sum_{i=0}^\nu \frac{\Delta M\ii}{M\ii}= J(k,r)\frac{\Delta r}{r}~~\\
\frac{\clM\tr(k,r)}{\clM}= \prod_{i=0}^{\nu} \frac{M'_{i+1}}{M\ii},~~~~~~~~~~~~~~~~~~~~~~~~~~~~~~~~~~~~~~~~~~
\label{finMp}
\eeqa
with $\tilde I_0(r)$ and $Y(k)$ being defined in equations (\ref{i}) and (\ref{Y}), respectively. 

\section{DF on the SUBHALO POPULATION}\label{population}

To analyse the effect of DF on the entire subhalo population in haloes with $M\h$ at $t\h$ we will follow the same strategy as in Papers II and III, that is, we will first focus on purely accreting haloes and then on ordinary ones having undergone major mergers. 

\subsection{Purely Accreting Haloes}\label{PA}

{\it In the absence of DF}, the inside-out growth of accreting haloes guaranties that the halo inside any radius $r$ stays unaltered. Consequently, all subhaloes accreted at $r$ follow the same fixed orbits, regardless of their mass and the effect of tidally stripped and shock-heated. In these conditions, the mean number of stripped subhaloes per infinitesimal truncated mass, $\clM\tr$, and radius $r$ within a halo with virial mass $M\h$ and virial radius $R\h$ at $t\h$ is given by (Paper II)
\beq
{\cal N}\fin(r,\clM\tr)
\!=\!\!\int_{0}^{v\maxi(r)}\!\!\! \der v \frac{\partial \clM}{\partial
  \clM\tr} {\cal
  N}\acc[v,r,\clM(v,r,\clM\tr)],
\label{corrbis} 
\eeq
where $v\maxi=\sqrt{GM(r)/r}$ is the maximum velocity at apocentre. In equation (\ref{corrbis}), ${\cal N}\acc(v,r,\clM)$ is the abundance of accreted subhaloes per infinitesimal mass $\clM$, apocentric radius $r$ and tangential velocity $v$, and $\partial \clM/\partial \clM\tr$ is the inverse Jacobian of the transformation $\clM\tr=\clM\tr(v,r,\clM)$ describing the stripping of accreted subhaloes. Strictly speaking, we should add a second term giving the abundance of stripped subhaloes arising from subsubhaloes released in the intra-halo medium from more massive stripped subhaloes. However, as shown in Paper II, this term contributes only to less than a few percent to the total abundance of stripped subhaloes at any radius $r$, so we can ignore it for simplicity.

Since the kinematics of accreted subhaloes does not depend on their mass, ${\cal N}\acc(v,r,\clM)$ factorises in the velocity distribution function, $f(v,r)$ (see Paper II for its form) times the mean abundance of accreted subhaloes (eq.~[17] of Paper I),
\beq
{\cal N}\acc(r,\clM)= 4\pi\,r^2 \frac{\rho(r)}{M\h}\,{\cal N}\acc(\clM).
\label{ratio}
\eeq
And, given that the MF of accreted subhaloes ${\cal N}\acc(\clM)$ is very nearly proportional to $\clM^{-2}$ (Paper I), equation (\ref{corrbis}) leads to
\beq
{\cal N}\fin(r,\clM\tr)
= \mu(r,\clM\tr)\, {\cal N}\acc(r,\clM\tr),
\label{firstterm} 
\eeq
where ${\cal N}\acc(r,\clM\tr)$ is the abundance of accreted subhaloes ending up with $\clM\tr$ and 
\beqa
\mu(r,\clM\tr)\!=\!\!\int_0^{v\maxi(r)}\!\! \der v \frac{f(v,r)(\clM\tr)^2}{\clM^2(v,r,\clM\tr)}\frac{\partial \clM(v,r,\clM\tr)}{\partial \clM\tr}~~~~\nonumber\\
\equiv \left\lav\frac{\partial \clM^{-1}}{\partial (\clM\tr)^{-1}}\right\rav(r,\clM\tr).~~~~~~~~~~~~~~
\label{mu} 
\eeqa
(with angular brackets denoting average over the subhalo velocities) is the truncated-to-original subhalo mass ratio at $r$ averaged over the velocity $v$ of accreted subhaloes at $r$, which is separable and very nearly a function of $r$ alone, $\mu(r)=\lav\clM\tr/\clM\rav(r)$. Its weak dependence on $\clM\tr$ (it is proportional to $(\clM\tr)^{-0.03}$) arises from the dependence of the subhalo concentration on mass (see Sec.~6 off Paper II). But, for simplicity, we adopt in what follows the approximation of a fixed subhalo concentration at accretion, equal to the typical concentration of haloes with masses $10^{-2}$ the mass $M(r)$ of the host halo at that moment, as done in Section 5 of Paper II. 

But, {\it in the presence of DF}, the preceding results do not hold because of the slight change from $\clM\tr$ to $\clM\ff$ and from $r$ to $r\ff$ of subhaloes. Replacing the abundance ${\cal N}\fin(r,\clM\tr)$ of stripped subhaloes per infinitesimal mass and radius around $\clM\tr$ at $r$ by the abundance ${\cal N}\rfin(r\ff,\clM\ff)$ of stripped subhaloes per infinitesimal final mass and radius around $\clM\ff$ and $r\ff$, the same derivation above then leads to 
\beqa 
{\cal N}\rfin(r\ff,\clM\ff)\!=\!\!\int_0^{v\maxi}\!\!\! \der v\left\{\frac{\partial \clM}{\partial
  \clM\ff}\frac{\partial r}{\partial
  r\ff}\!+\! \frac{\partial \clM}{\partial
  r\ff} \frac{\partial r}{\partial
  \clM\ff}\!\right\}\!(v,r,\clM)\nonumber\\ \times\,{\cal
  N}\acc(v,r,\clM),~~~
\label{corrbisnew} 
\eeqa
where $v\maxi$ is the solution $v$ of the implicit equation $v=\{GM[r(v,r\ff,\clM\ff)]/r(v,r\ff,\clM\ff)\}^{1/2}$. For simplicity in the notation, we have omitted in the right-hand member of equation (\ref{corrbisnew}) the explicit dependence of $v\maxi$, $r$ and $\clM$ on $r\ff$ and $\clM\ff$. Again, the relation (\ref{corrbisnew}) can be rewritten in the form
\beq 
{\cal N}\rfin(r\ff,\clM\ff)=\mu\DF(r\ff,\clM\ff)\, {\cal
  N}\acc(r\ff,\clM\ff),
\label{firsttermnew} 
\eeq
where
\beqa
\mu\DF(r\ff,\clM\ff)=\int_0^{v\maxi}\!\der v f[v,r)\,\frac{r^2\rho(r)}{(r\f)^2\rho(r\ff)}\left(\frac{\clM\ff}{\clM}\right)^{\!\!2}(v,r,\clM)\!\!\!\!\!\!\!\!\nonumber\\
\times\left\{\frac{\partial \clM}{\partial
  \clM\ff}\frac{\partial r}{\partial
  r\ff}\!+\! 
  \frac{\partial \clM}{\partial
  r\ff} \frac{\partial r}{\partial\clM\ff}\right\}(v,r,\clM)\!\!\!\!\!\!\!\!
\nonumber\\
\equiv\left\lav\frac{\partial \clM^{-1}}{\partial
 (\clM\ff)^{-1}}\frac{\partial M(r)}{\partial
  M(r\ff)}\!+\! 
  \frac{\partial \clM^{-1}}{\partial
  M(r\ff)} \frac{\partial M(r)}{\partial
  (\clM\ff)^{-1}}\right\rav
\label{munew} 
\eeqa
(with angular brackets being again the average over the velocity $v$ at $r\ff$; see App.~\ref{App3}). When DF is negligible so that $r\ff$ equals $r$ and $\clM\ff$ equals $\clM\tr$, ${\cal N}\fin$ becomes ${\cal N}\rfin$ and $\mu\DF(r\ff,\clM\ff)$ becomes $\mu(r)$ (eq.~[\ref{mu}]), so the new expression (\ref{mu}) holds in general, regardless of how strong is DF. 

In Appendix \ref{App3} we derive the explicit form of $\mu\DF(r\ff,\clM\ff)$ to first order in $\Delta E/E$ and $\Delta L/L$. The result is
\beqa
\mu\DF(r\ff,\clM\ff)
\approx\mu(r\ff)\left[1+\omega(r\ff)\clM\ff\right]~~~~~~~~~~~~~~~~~~~~~~~~~\label{muDF2}\\
\omega(r\ff)\equiv \frac{\kappa(r\ff)}{\mu(r\ff)}\frac{\der\ln (M\tilde I_0)}{\der\ln r\ff}\tilde I_0(r\ff),~~~~~~~~~~~~~~~~~~~~~~~~~~~~~~~\label{omega}
\eeqa
%
where $\kappa(r\ff)$ is defined as the average over $v$ of $Y(k)\clM\tr/\clM$. Equation (\ref{muDF2}) states that $\mu\DF$ is equal to its counterpart in the absence of DF plus one positive first order term dependent on the subhalo mass. Thus, while $\mu$ is a function of the radius only, $\mu\DF$ also depends on subhalo mass, as expected.  

\begin{figure}
\centerline{\includegraphics[scale=0.465, bb= 280 31 300 560]{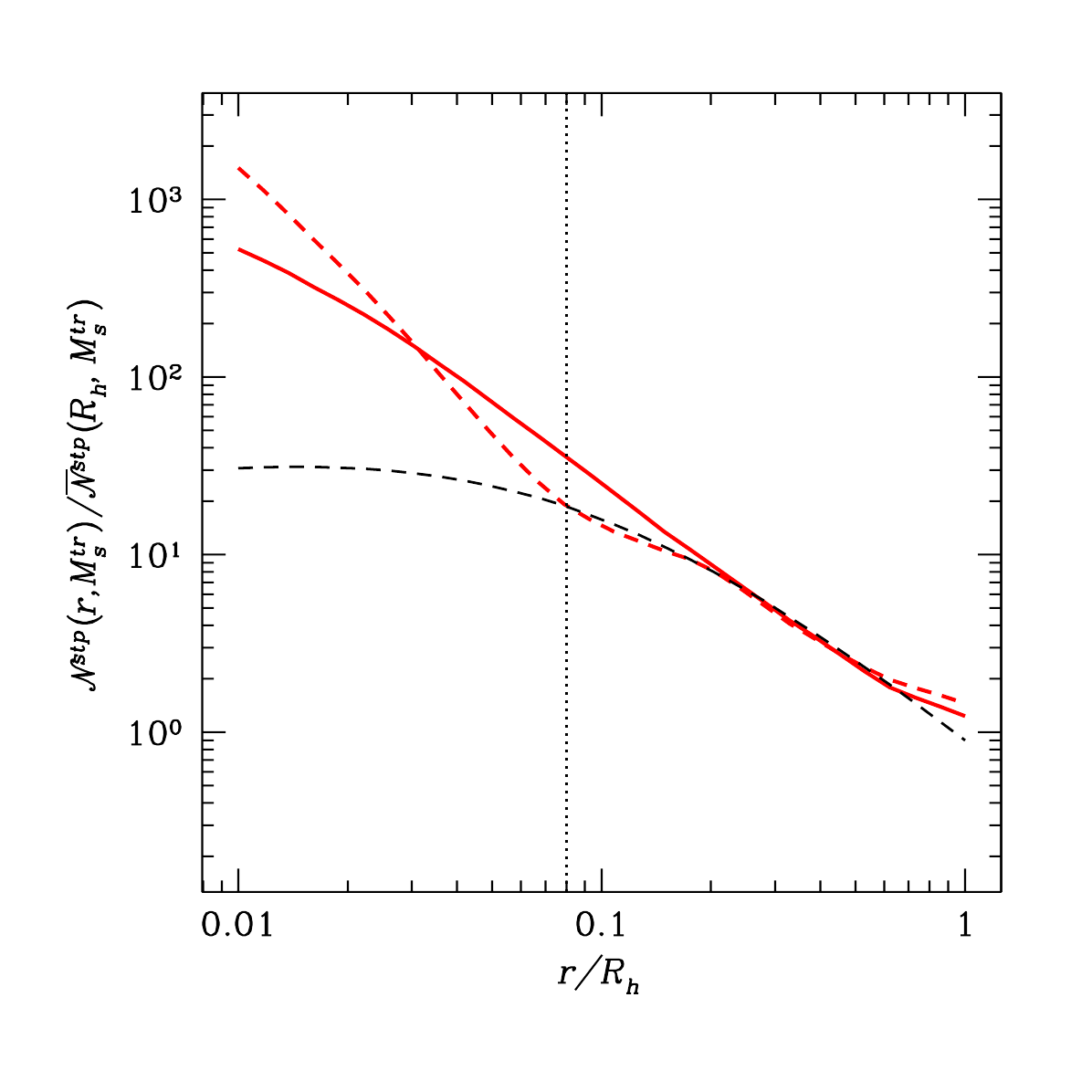}}
\caption{Radial abundance of stripped subhaloes of mass $\clM\tr$ scaled to their total abundance (the scaled quantity is independent of subhalo mass) in current purely accreting MW-mass haloes calculated using the $M$-$c$ relations derived by \citet{Sea23} using CUSP (solid red line) and found by \citet{Gea08} (dashed red line) in the Millennium simulation with a limited halo mass resolution. Both solutions are essentially proportional to $r^{1.3}\rho(r)$ (dashed black line) at radii larger than $r/R\h= 0.08$ (vertical dotted line) as found by \citet{Hea16} in the MW-mass AqA1 halo in that simulation.\\
(A colour version of this Figure is available in the online journal.)}
\label{fnew}
\end{figure}

\begin{figure}
\centerline{\includegraphics[scale=1.25, bb= -20 0 290 200]{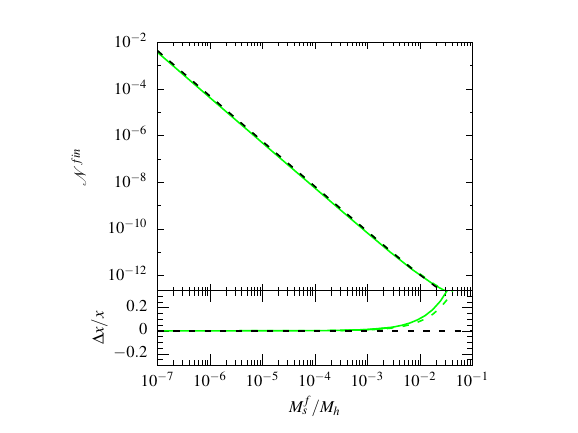}}
\caption{Differential subhalo MF with DF in current purely accreting MW-mass haloes in the {\it WMAP7} cosmology derived using the CUSP (solid green line) and Gao et al.'s (dashed green line) $M$-$c$ relation, compared to the predictions without DF (dashed black line).\\
(A colour version of this Figure is available in the online journal.)}
\label{f2}
\end{figure}

Having determined $\mu\DF$, the radial abundance of evolved subhaloes of mass $\clM$ takes the form 
\beq
{\cal N}\rfin(r\ff,\clM\ff)\approx{\cal N}\fin(r\ff,\clM\ff)\left[1+\omega(r\ff)\clM\ff\right],
\label{calN}
\eeq
showing that the subhalo radial abundance in the presence of DF increases inwards compared to that with no DF through the function $\omega(r\ff)$, the difference being proportional to $\clM\ff$. Factor $\omega(r)$ is roughly proportional to $M\h^{-1}$ (through factor $\rho(r)/\sigma^3(r)$ in the function $\tilde I_0$; see App.~\ref{App3}), so the effect of DF on subhaloes with $\clM\ff/M\h$ is similar for haloes of all masses, just slightly less marked in more massive ones.

In what follows, to illustrate our predictions we use the abundance of stripped subhaloes without DF, ${\cal N}\fin(r,\clM\tr)$, calculated using: 1) the $M$-$c$ relation found by \citet{Gea08} in the Millennium simulation, affected by a limited halo mass resolution, and 2) the $M$-$c$ relation derived in \citet{Sea23} by means of CUSP, free of this effect. Indeed, as discussed in Paper II, the concentration of the halo plays a crucial role on tidal stripping and shock-heating, so the ${\cal N}\fin(r,\clM\tr)$ markedly depends on whether the $M$-$c$ used has been derived with or without a limited halo mass resolution altering it at low-masses and high-$z$. We see in Figure \ref{fnew} that the resulting ${\cal N}\fin$ are quite different at radii smaller than $r/R\h\sim 0.08$, indeed, because stripping is very sensitive to the concentration of haloes (see Paper II). As a consequence, the radial abundance of stripped subhaloes derived using Gao et al.'s $M$-$c$ relation increases much more steeply inwards than that derived using CUSP. However, both profiles reproduce the profile of low-mass subhaloes found by \citet{Hea16} at $r/R\h>0.08$ in the purely accreting MW-mass AqA1 halo in the Millennium simulation, except for a small edge effect at $r\sim R\h$ due to the approximate procedure used to find the properties of subhaloes at accretion (Sec.~\ref{multiple}), which is hereafter corrected in order to avoid any spurious effect. 

Integrating the radial abundance (\ref{calN}) over $r\ff$, we obtain the differential subhalo MF, which takes the form
\beq
{\cal N}\rfin(\clM\ff)\approx{\cal N}\fin(\clM\ff)\left[1+\bar\omega(R\h)\clM\ff\right].
\label{MF}
\eeq
From now on, a bar on a function of radius denotes the stripped subhalo number-weighted average of that function out to that radius. Equation (\ref{MF}) shows that the subhalo MF with DF differs from that without through a term proportional to $\clM$, with constant $\bar \omega(R\h)$. This constant is very nearly proportional to $M\h^{-1}$, so the relative difference between the subhalo MFs with and without DF is essentially the same for all halo masses. 

In Figure \ref{f2} we show the predicted subhalo MF resulting for MW-mass haloes using the two different $M$-$c$ relations mentioned above, the result being very similar in both cases. In the upper panel we see that DF has no apparent effect on the subhalo MF at the scale usually used to estimate its logarithmic slope. It is thus unsurprising that the MFs derived in Papers II and III with no DF agreed so well with those found in simulations. In the lower panel we see, however, that the relative difference with respect to th MF without DF increases with increasing mass from a few percent at $\clM\ff\sim 10^{-3}M\h$ to $\sim 35$\% at the maximum subhalo mass, $\clM\ff\sim \bar\mu(R\h) 0.3\times 10^{-1} M\h\sim 0.03M\h$ (captured haloes more massive than $M\h/3$ give rise to major mergers, not to subhaloes). Nevertheless, the difference is moderate because the effect of DF in the strength of tidal stripping over one orbit is of second order only (see App.~\ref{App3}). The first order effect is due to the drift of massive subhaloes inwards, causing their tidal stripping to be less effective (Paper II). This is why the number of massive evolved subhaloes increases with respect to the case without DF. 

\begin{figure}
\centerline{\includegraphics[scale=1.18, bb= -5 -6 290 205]{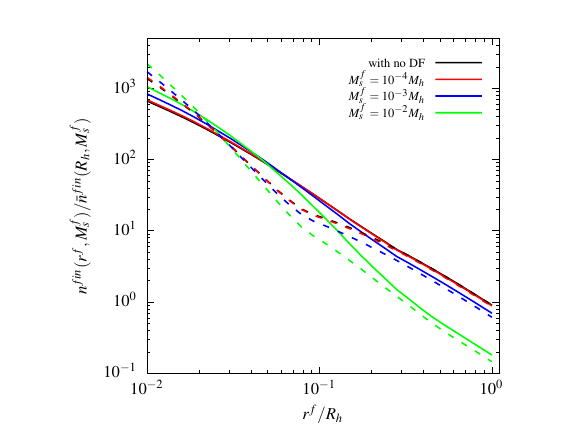}}
\caption{Scaled number density profiles of subhaloes of several masses in the same haloes as in Figure \ref{f2} (coloured lines) derived using the CUSP (solid lines) and Gao et al.'s (dashed lines) $M$-$c$ relations, compared to the profiles found without DF, which overlap in one single curve (black line) and with the profiles with DF of subhaloes of $\clM\ff\la  10^{-4}M\h$.\\
(A colour version of this Figure is available in the online journal.)}
\label{f3}
\end{figure}

On the other hand, scaling the number density profile $n\rfin(r\ff,\clM\ff)={\cal N}\rfin(r\ff,\clM\ff)/[4\pi (r\ff)^2]$ of subhaloes with mass $\clM\ff$ given above to the total number density of such subhaloes $\bar n\rfin(R\h,\clM\ff)=3{\cal N}\rfin(\clM\ff)/[4\pi R\h^3]$, we obtain the expression
\beq
\frac{n\rfin(r\ff,\clM\ff)}{\bar n\rfin(R\h,\clM\ff)}\!\approx\!\frac{n\fin(r\ff,\clM\ff)}{\bar n\fin(R\h,\clM\ff)}\!\left\{1\!+\!\left[\omega(r\ff)\!-\!\bar\omega(R\h)\right]\clM\right\}\!.
\label{newn}
\eeq
Again, the scaled subhalo number density profile is equal to its counterpart in the absence of DF, independent of subhalo mass, plus a term proportional to $\clM\ff$, with proportionality factor $\omega(r\ff)-\bar\omega(R\h)$. Since $\omega(r\ff)$ is positive and decreases with increasing radius, factor $\omega(r\ff)-\bar\omega(R\h)$ starts being positive at small radii and ends up being negative at larger ones, causing the profiles to be steeper than without DF. As can be seen in Figure \ref{f3}, contrary to what happened with the subhalo MF the scaled subhalo number density profiles depend substantially on the $M$-$c$ relations used to derive them though they always show the same expected behaviour: they become increasingly steep for subhaloes more massive than $10^{-4}M\h$ and overlap with each other and with the profiles without DF for less massive ones.

\begin{figure}
\centerline{\includegraphics[scale=1.18, bb= -5 -6 290 205]{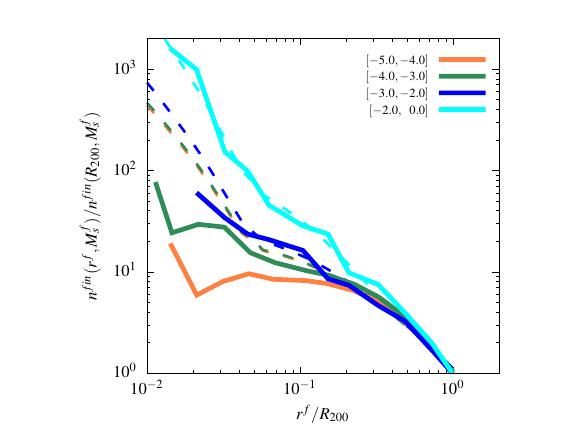}} 
\caption{Number density profiles normalised to $R_{200}$ of subhaloes with masses $\clM/M_{200}$ in the quoted bins in haloes at $z=0$ with $M_{200}$ in the range $[10^{13}h^{-1}$\modotc,$10^{14}h^{-1}$\modotc] found by \citet{Hea18} in the Millennium simulation (broken coloured thick lines), compared to our predictions for subhaloes with $\clM/M_{200}$ equal to the lower bounds of those bins, by far the most numerous in each bin, in current purely accreting haloes with $M_{200}=0.5\times 10^{14}h^{-1}$ \modot and concentration $c_{200}=6.5$ (dashed coloured lines).\\
(A colour version of this Figure is available in the online journal.)}
\label{comparison}
\end{figure}

But to better assess the goodness of the model we must compare our predictions to simulations. For this comparison, we have used the subhalo number density profiles in current haloes obtained by \citet{Hea18} in the Millennium simulation. Even though the formation time of such haloes is not specified, since the earlier haloes form, the more spherical and smoother they are and the better their subhalo number density profiles can be determined, the sample should be strongly biased to early forming objects, so the comparison with purely accreting haloes should be compelling. 

As can be seen in Figure \ref{comparison},\footnote{The mass $M_{200}$ used is the mass inside the radius $R_{200}$ encompassing an inner mean density equal to 200 times the critical density.} our predictions fully reproduce the numerical data within their uncertainty\footnote{The error bars are unknown, but they can be guessed from the oscillations of the broken lines.} down to $r/R_{200}\sim 0.1$ in general and even beyond for massive subhaloes. There is just a trend for the profile of the less massive subhaloes in simulations to be shallower than predicted, a trend which becomes increasingly marked and affects increasingly massive subhaloes with decreasing radius. This effect is obviously not due to DF as it affects subhaloes less massive than $10^{-4}M_{200}$. As mentioned by \citet{Hea18}, there is indeed a depletion of subhaloes at the central parts of haloes because, when objects with radially elongated orbits ($k\ll 1$) and small enough velocities (as found at small radii) pass through the central dark matter condensation of the halo, they merge with it and disappear. This effect is more apparent for low-mass subhaloes simply because, contrarily to what happens with massive subhaloes subject to DF, the steep increase in their abundance at small radii due to stripping (with Gao et al.'s $M$-$c$ relation; see Fig.~\ref{fnew} for MW-mass haloes) is insufficient to hide, in logarithmic scale, their depletion. Nevertheless, at small enough radii, that steep increase overcomes the increasing depletion and the abundance of evolved subhaloes of any mass rapidly increases again. The possible merging of subhaloes with the central dark matter condensation of haloes was not taken into account in Paper II when deriving the radial abundance of stripped subhaloes ${\cal N}\fin(r,\clM\tr)$ without DF (and comparing it to that found by \citet{Hea16} in the simulated AqA1 halo at $r/R\h > 0.08$), so we cannot expect to recover it now. But what is important to retain from this comparison is that the modelled effect of DF has the right overall behaviour and the right specific amplitude dependent on subhalo mass found in simulations.


\subsection{Ordinary Haloes}\label{ordinary}

But haloes alternate accretion periods with major mergers and, even though all halo properties arising from gravitational collapse and virialisation do not depend on their past aggregation history \citep{SM19}, those linked to tidal stripping and shock-heating {\it and} DF do. Thus, the properties of substructure depend not only on their halo mass $M\h$ and time $t\h$, but also on their formation time $t_{\rm f}$, defined, like in previous Papers, as the time they suffered the last major merger. Fortunately, as shown in Paper III, the properties of substructure in ordinary haloes can be derived from those in idealised purely accreting ones. Below, we focus on these properties {\it averaged over the halo formation time} and indicate how to calculate those of haloes formed in specific time intervals. To distinguish the properties of purely accreting haloes from those of ordinary ones the former, derived in Section \ref{PA}, are hereafter denoted with index PA.  

When a halo undergoes a major merger, it virialises through violent relaxation, which `scrambles' its content (i.e. the location and velocity of subhaloes in the new halo are randomly reshuffled across the halo keeping the constraints of its final equilibrium configuration). After that, it begins to accrete again and to grow inside-out. As a consequence, the mean fraction of accreted subhaloes with $\clM\ff$ at $r\ff$ in ordinary haloes with $M\h$ at $t\h$ that are stripped is (Paper III)
\beqa
{\mu\DF}{}_{\rm [M\h,t\h]}(r\ff,\clM\ff)=\int_0^{t(r\ff)} \der \tilde t \,n(\tilde t)\,\mu\DF\PA{}_{\rm [M\h,t\h]}(r\ff,\clM\ff)~~~\nonumber\\
+  \int_{t(r\ff)}^{t\h} \der \tilde t\, n(\tilde t)\, 
\bar\mu\DF{}_{\rm [M(\tilde t),\tilde t]}(R\h,\clM\ff).~~~
\label{one2}
\eeqa
In equation (\ref{one2}), $n(t)$ is the formation time probability distribution function (PDF) of haloes with $M\h$ at $t\h$ calculated by \citet{Mea98} using CUSP (see \citealt{Rea01} for a practical approximation). Since $\mu\DF$ of haloes with $M(t)$ at $t$ depends on these two quantities, we write it with subindex [M(t),t]. Consequently, the integral over $t$ in the first term on the right of equation (\ref{one2}) can be rewritten as $\mu\PA\DF{}_{\rm [M\h,t\h]}(r\ff,\clM\ff)\,n\uc(r\ff)$, where $n\uc(r\ff)$ is the cumulative formation time PDF of haloes with $M\h$ at $t\h$ up to the time $t(r\ff)$ when they reached radius $r\ff$ and mass $M(r\ff)$. 

Following Paper III, the relation (\ref{one2}) can be used to obtain $\bar \mu\DF$ from the homologous function $\mu\DF\PA$ derived in Section \ref{PA}. To do this we must first multiply it by ${\cal N}\acc(r\ff,\clM\ff)$ and integrate over $r\ff$ out to $R\h$. The result is
\beqa
\bar\mu\DF{}_{\rm [M\h,t\h]}(R\h)=\overline{\,n\uc\mu\DF\PA{}_{\rm [M\h,t\h]}}(R\h)~~~~~~~~~~~~~~~~~~~~~~~~~~\nonumber\\
\!+\!\!\int_0^{t\h}\!\! \der \tilde t\, n(\tilde t)\,
\bar\mu\DF{}_{\rm [M(\tilde t),\tilde t]}(R\h)\frac{M(\tilde t)}{M\h},~~~~~
\label{one3}
\eeqa
where $\bar\mu\DF{}_{\rm [M(\tilde t),\tilde t]}$ appears to be independence of $\clM$. The first term on the right of equation (\ref{one3}) is related to its counterpart with no DF (eq.~[\ref{muDF2}]) through 
\beqa
\overline{n\uc{\mu\DF\PA}_{\rm [M\h,t\h]}}(r\ff,\clM\ff)=
\overline { n\uc{\mu\PA}_{\rm [M\h,t\h]}}(r\ff)
\left[1+\widetilde\omega(r\ff)\clM\ff\right]~\\
\widetilde\omega(r\ff)\equiv\frac{\overline{n\uc\mu\PA_{\rm [M\h,t\h]}\omega\PA(r\ff}}{\overline {n\uc\mu\PA_{\rm [M\h,t\h]}}(r\ff)}.\,~~~~~~~~~~~~~~~~~~~~~~~~~~~~~~~~~~~~~~
\eeqa
Solving the Volterra equation (\ref{one3}) for ${\bar\mu\DF}{}_{\rm [M(t),t]}(R\h)$ (and the analogous Volterra equation for ${\bar\mu}{}_{\rm [M(t),t]}(R\h)$) and replacing it into equation (\ref{one2}), we arrive after some algebra at 
\beqa
{\mu\DF}_{\rm [M\h,t\h]}(r\ff,\clM\ff)={\mu}_{\rm [M\h,t\h]}(r\ff)
\left[1+\omega(r\ff)\clM\ff\right]~~~~~~~~~~\label{muord}\\
\omega(r\ff)\!\equiv\! \frac{\{n\uc\mu\PA_{\rm [M\h,t\h]}\omega\PA\}(r\ff)}{{\mu}_{\rm [M\h,t\h]}(r\ff)}~~~~~~~~~~~~~~~~~~~~~~~~~~~~~~~~~~~~\nonumber\\
+\frac{\int_{t(r\ff)}^{t\h}\!\der \tilde t \,n(\tilde t)\,{\bar\mu\DF}{}_{\rm [M(\tilde t),\tilde t]}(R\h)\!-\!{\bar\mu}{}_{\rm [M(\tilde t),\tilde t]}(R\h)}{{\mu}_{\rm [M\h,t\h]}(r\ff)}.~~~~~~\label{newomega}
\eeqa
Thus, $\mu\DF$ of ordinary haloes with $M\h$ at $t\h$ has the same form as for purely accreting haloes (eq.~[\ref{muDF2}]), but with a different $\omega(r\ff)$, related to its counterpart in pure accretion, $\omega\PA(r\ff)$, through equation (\ref{newomega}).

\begin{figure}
\centerline{\includegraphics[scale=1.18, bb= -5 -7 290 205]{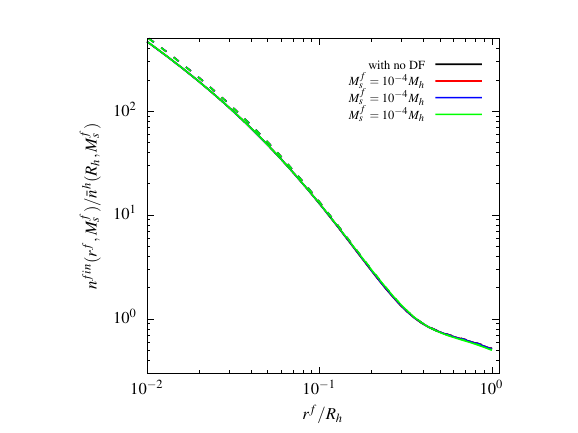}}
\caption{Same profiles as in Figure \ref{f3} but averaged over halo formation times.\\
(A colour version of this Figure is available in the online journal.)}
\label{f4}
\end{figure}

Having determined $\mu\DF{}_{\rm [M\h,t\h]}(r\ff,\clM\ff)$, we must plug it in equation (\ref{firsttermnew}) in order to obtain the formation time-averaged abundance of evolved subhaloes having suffered stripping and DF, ${\cal N}\rfin(r\ff,\clM\ff)$. Then, integrating the latter quantity over $r\ff$, we are led to the formation time-averaged subhalo MF in ordinary haloes with DF and, following the same procedure as for purely accreting haloes, to their scaled subhalo number density profiles. Of course, since the subhalo MF of purely accreting haloes is almost identical to that found without DF, so should also be the subhalo MF of ordinary haloes with DF derived from it. On the contrary, since the scaled subhalo number density profile with DF in purely accreting haloes depends on subhalo mass, we might expect the same behaviour in ordinary haloes averaged over their formation times. However, as shown in Figure \ref{f4}, the scrambling produced in haloes with different formation times fully erases the difference between subhaloes of distinct masses and even between the distinct $M$-$c$ relation used to derive them.

All previous results referred to ordinary haloes of fixed mass averaged over all their formation times. To obtain the subhalo MF and radial distribution in ordinary haloes formed in specific time intervals we should simply apply the same procedure, but with the formation time PDF, $n(t)$, multiplied by the suited top-hat window defining the desired time interval (see Paper III). In this case, the effects of DF would be substantially more marked for haloes formed long time ago than formed recently because DF would have had more time to proceed. Specifically, the earlier haloes would form, the closer their properties would be to those of purely accreting haloes.

\section{Summary and Concluding Remarks}\label{sum}

With this Paper we culminate a detailed comprehensive study of halo substructure. This has been accomplished in a fully analytic manner in the peak model of structure formation (with no free parameter) together with a realistic model of subhalo tidal stripping and shock-heating (with only two parameters tuned by means of numerical experiments). 

In Paper I we derived the properties of unevolved subhaloes falling into purely accreting haloes, using the statistics of nested peaks. In Paper II we studied their tidal stripping and shock-heating within the host haloes. And in Paper III we extended the analysis to ordinary haloes of all masses and formation times. The theoretical properties obtained accurately reproduced those found in cosmological $N$-body simulations and explained them. However, to facilitate the analytic treatment we neglected the effects of DF, so the results held for low-mass subhaloes only. 

In the present Paper, we have remedied that limitation by incorporating DF, taking into accounts its cross-effects with tidal stripping and shock-heating of subhaloes. After monitoring the multiple concatenated orbits of individual subhaloes at accretion and finding their final mass and radius at the time haloes are observed, we have studied the effects of DF on their global subhalo population in both purely accreting haloes and ordinary ones having undergone major mergers. 

As expected, we have found that the orbital decay by DF of massive subhaloes causes their number density profiles to become notably steeper than without DF. Specifically, the predicted profiles are in good agreement with the results of simulations provided the effects of their limited halo mass resolution is taken into account. The drift of massive subhaloes toward small radii also causes them to be less stripped, which tends to increase their abundance with respect to the case without DF. Nevertheless, the  change in the subhalo MF is insignificant at the scale usually use to estimated its logarithmic slope. This explains why the subhalo MF predicted in Papers II and III agreed so well with that found in simulations affected by DF.

Despite the entanglement of the different mechanisms driving the properties of halo substructure, we have obtained simple non-parametric expressions for the subhalo MF and number density profiles showing how DF alters the results obtained in previous Papers. These expressions not only explain the results of high-resolution $N$-body simulations, correcting them for the effects of the limited halo mass resolution, but also extend those results to haloes of any mass $M\h$, redshift $z$ and formation time $t$ in any desired CDM cosmology. Thus they stand as a useful tool for cosmological studies.

To conclude we would like to mention possible applications of the present work. Given the correlation between galaxy stellar mass and the mass of their subhalo hosts, the results given here open the possibility to determine the galaxy bias directly from peak statistics without the need to model the halo occupation distribution of galaxies. In addition, with small modifications, the present analytic treatment of DF should also be possible to apply to the modelling of other self-gravitating systems.



\par\vspace{0.75cm}\noindent
{\bf DATA AVAILABILITY}

\vspace{11pt}\noindent 
The numerical codes used in the CUSP formalism and the present series of Papers are available from \url{https://gitlab.com/cosmoub/cusp}.

\vspace{0.75cm} \par\noindent
{\bf ACKNOWLEDGEMENTS} 

\vspace{11pt}\noindent
This work was funded by the Spanish MCIN/AEI/10.13039/ 501100011033 through grants CEX2019-000918-M (Unidad de Excelencia `Mar\'ia de Maeztu', ICCUB) and PID2022-140871NB-C22 (co-funded by FEDER funds) and by the Catalan DEC through the grant 2021SGR00679 .

{}


\appendix

\onecolumn

\section{Relative Energy and Angular momentum Increments}\label{App1}

To leading order in the effects of DF, Equation (\ref{Linvariant}) implies
\beq
\frac{\Delta L}{L}= -A\LL(k,r,\clM) T(k,r,\clM),
\label{AL}
\eeq
where $T(k,r,\clM)$ is the orbital period with DF, equal, to leading order, to that with no DF. On the other hand, equation (\ref{Einvariant}) states that $\Delta E$ is $-A\E$ times the action of the mechanical system, for slowly varying $E$, is an adiabatic invariant, so it is equal to the value found in the absence of DF. Consequently, equation (\ref{Einvariant}) implies
\beq
\frac{\Delta E}{E}=P(k,r) A\E(k,r,\clM)  T(k,r,\clM),
\label{AE}
\eeq
where
\beq
P(k,r)\equiv \frac{-2r}{kGM(r)S(k,r)}\frac{\int_0^{T} \der t v^2(t)}{\int_0^{T} dt}= \frac{-2r}{kGM(r)S(k,r)}\frac{\int_{r\per}^{r}\der x \frac{v^2(x,k,r)}{v_{\rm r}(x,k,r)}}{\int_{r\per}^{r}\der x \frac{1}{v_{\rm r}(x,k,r)}},
\label{P}
\eeq
with the second equality on the right holding to leading order in the effects of DF, where
\beq
v^2(x,k,r)=2[\Phi(r)-\Phi(x)]+k\frac{GM(r)}{r}~~~~~~~~~~~~~~~{\rm and}~~~~~~~~~~~~~~~~~
v^2_{\rm r}(x,k,r)=2[\Phi(r)-\Phi(x)]+k\frac{GM(r)}{r}\left[1-\left(\frac{r}{x}\right)^2\right]
\label{vs}
\eeq
(with $0<(v-v_{\rm r})/v<1$) are the 3D velocity and its radial component, respectively, at radii $x$ over the orbit without DF of subhaloes with $k$ and $r$ and $r\per/r\equiv \tilde k\approx k(1+\sqrt{1+8/k})/4$.\footnote{This relation results from energy and angular momentum conservation in the orbit without DF, Taylor expanding to first order $\Phi(r\per)$ around $\Phi(r)$.} 

The functions $P A\E T$ and $A\LL T$ determining $\Delta E/E$ and $\Delta L/L$ (eqs.~[\ref{AE}] and [\ref{AL}]) can be calculated, to leading order, from equations (\ref{Einvariant}) and (\ref{Linvariant}), with $A$ given by equation (\ref{A1}), over orbits without DF, i.e. using $v$ and $v_{\rm r}$ given by equations (\ref{vs}). The result is
\beq
\frac{\Delta E}{E}=P(k,r)A\E(k,r,\clM) T(k,r,\clM) 
= 
30.86\,\pi G^2 Q \frac{\clM}{M\h}\,\frac{-2r}{kGM(r)S(k,r)}
\int_{r\per}^{r} \der x \,x^{-1.875}f\dc(x)\frac{v^2(x,k,r)}{v_{\rm r}(x,k,r)}H(x,k,r)
\label{AE2}
\eeq
\beq
\frac{\Delta L}{L}=-A\LL(k,r,\clM) T(k,r,\clM) 
= -30.86\,\pi G^2 Q \frac{\clM}{M\h}\int_{r\per}^{r} \der x\,x^{-1.875}f\dc(x)\frac{1}{v_{\rm r}(x,k,r)}H(x,k,r),
\label{AL2}
\eeq
where $H(x,k,r)\equiv [{\rm erf}(X)-\sqrt{2}/\pi X\exp(-X^2)]/X^3$, with $X\equiv\sqrt{3/2}\,v(x,k,r)/\sigma (x)$. To write equations (\ref{AE2}) and (\ref{AL2}) we have used the universal dDM fraction $f\dc(r)$ (Papers II and III) and the pseudo phase-space density $\rho(r)/\sigma^3(r)=Q/M\h\, r^{-1.875}$, with $Q=9.56\times 10^{17}$ (M$_\odot$/Mpc$^3$) (km/s)$^{-3}$ (\citealt{TN01,SM19} and references therein). The case $k=1$ ($r\per=r$) is excluded from expressions (\ref{AE2}) and (\ref{AL2}). But this is not a problem because this corresponds to the extreme value of $k$ for accreted subhaloes (Paper II).

These values of $\Delta E/E$ and $\Delta L/L$ can be calculated with no need to previously determine the subhalo orbits with DF. However, they still involve two numerical integrals. To get a purely analytic treatment we can split those integrals in two parts and Taylor expand to first order the integrands around the lower and upper radii, which allows us to carry out the integrals analytically. The separation radius $r\cc$ is taken such that the approximate values of $v_{\rm r}$, the most sensitive quantity (it is a first order term at the denominator of the integrants; see below), coincide at that matching radius of both parts. This leads to $r\cc(k,r)/r\per-1$ equal to the minimum positive (or null) solution of the quadratic equation $Ax^2+Bx+C=0$, with
\beq
A=q_u(k,r)[\tilde k^4\!-\!q_l(k.r)],~~B=1-\frac{q_l(k,r)}{k}+\tilde k^3\left[\frac{1}{k}\!-\!1\!-\!2(1\!-\!\tilde k)q_u(k,r)\right]~~ {\rm and}~~ C=-\tilde k^2(1\!-\!\tilde k)\left[\frac{1}{k}\!-\!1\!-(1\!-\!\tilde k)\!q_u(k,r)\right],
\label{eq}
\eeq
being $q_u(k,r)\equiv (\der\ln M/\der\ln r-2)/k+3$ and $q_l(k,r)\equiv \tilde kM(\tilde k r)/M(r)$. We remark that 
$q_u(k,r)$ and $q_l(k,r)$ are weakly dependent on $r$\footnote{Under the approximation $\rho(r) \propto r^{-2}$, we have $q_u(k)\approx -1/k+3$ and, expanding $M(\tilde k r)$ around $M(r)$ up to first order, we obtain $q_l(k) \approx \tilde k^2$.} and so is also $\tilde k\cc=r\cc/r\per$.

In the upper radial parts, we have (see eqs.~[\ref{vs}])
\beqa
v^2(x,k,r)\approx k\frac{GM(r)}{r}\left[1-\frac{2}{k}\left(\frac{x}{r}-1\right)\right],
\quad\quad~ v^2_{\rm r}(x,k,r)
\approx 2v^2(x,k,r)\left(1-\frac{x}{r}\right)\left[\frac{1}{k}-1+\left\{2+\frac{\der\ln [M(r)/r^2]}{\der \ln r}\right\}\left(\frac{x}{r}-1\right)\right],~
\\ f\dc(x)\approx f\dc(r)\left[1+\frac{\der \ln f\dc}{\der \ln r}\left(\frac{x}{r}-1\right)\right]~~~~~~~~~~~~~~~~~~~~~~~~~~~~~~~~~~~~~~~~~~~~~~~~~~~~~~~~~~~~~~~~~~~~~~~~~~~~~~~~~~~~~~~~~~~~~~~~~~~~~~
\eeqa
and, given the isotropic Jeans equation and the form of $\rho(r)/\sigma^3(r)$, 
\beq
\frac{2\sigma^2(x)}{3v^2(x,k,r)}\frac{\der \ln \left[\rho^{5/3}(x)\, x^{1.25}\right]}{\der \ln x}\!=\!-\frac{2 G M(x)}{xv^2(x,k,r)}\!\approx\! \frac{-1}{k/2-(x/r-1)}\left\{1+\frac{\der\ln [M(r)/r]}{\der \ln r}\left(\frac{x}{r}-1\right)\right\},
\eeq
(with $\rho(r)\sim r^{-2.5}$ at the radii of interest) implies $3v^2(r)/[2\sigma^2(r)]\approx 1.46 k$,
\beq
H(x,k,r)\approx  H(k)\left[1-\frac{\partial\ln H}{\partial\ln x}\Big|_{r}\left(\frac{x}{r}-1\right)\right],~~~~~~~~~~~~~~~{\rm with}~~~~~~~~~~~~~~~~\frac{\partial \ln H}{\partial \ln x}\Big|_{r}\approx \frac{\der\ln H(k)}{\der\ln k} \left[\frac{2}{k} + \frac{\der \ln M(r)/r}{\der \ln r} \right]
\eeq
and $H(k)\equiv (1.46k)^{-3/2}[{\rm erf}(\sqrt{1.46k})-2\sqrt{1.46k/\pi}\,\exp(-1.46k)]$.

In these conditions, the integrals in equations (\ref{AE2}) and (\ref{AL2}) above $r\cc$ take the fully analytic form
\beq
\frac{\Delta E}{E}\bigg|_{u}\!\approx  -\frac{2\clM I_0(r) H(k)}{S(k,r)}\left(\!\frac{1-\tilde k}{1\!-\!k}\!\right)^{\!\frac {1}{2}}I_{1}\!\Big[\tilde k,D(k,r,1)-\frac{2}{k}\Big]~~~~~~~~~~
\frac{\Delta L}{L}\bigg|_{u}\!\approx  -\clM I_0(r) H(k)\left(\!\frac{1-\tilde k}{1-k}\!\right)^{\!\frac{1}{2}}I_{1}\!\left[\tilde k, D(k,r,1)\right],~
\eeq
where
\beqa
I_0(r)\equiv 43.64\pi G^2f\dc(r)\frac{\rho(r)}{\sigma^3(r)} r\left[\frac{GM(r)}{r}\right]^{-{1/2}}~~~~~~~~~~~~~~
I_{1}(\tilde k,D)\equiv 1+0.625(1-\tilde k)\left\{1-\left[\frac{1}{1.875}+\frac{3(1+\tilde k)}{5}\right]D\right\},~~~~~~~~~~~~~~~~~\label{I0}\nonumber\\
{\rm with}~D(k,r,q)\equiv\frac{\der\ln f\dc}{\der\ln r}-\frac{\partial \ln H}{\partial \ln x}\Big|_{r}- \!\frac{q}{2(q-k)}\frac{\der\ln [M(r)/r^{2-3k}]}{\der\ln r},~{\rm so}~I_1(\tilde k,D)~{\rm is~weakly~dependent~on~} r~{\rm provided~} q~{\rm is}.\,~~~~~~~~~~~~~\nonumber
\eeqa

And proceeding in a similar way, the integrals in the lower radial parts lead to
\beq
\frac{\Delta E}{E}\bigg|_{l}\!\approx  -\frac{2\clM  I_0(\tilde k r)H(k)}{S(k,\tilde k r)}\left(\!\frac{1-\tilde k\cc}{q_{l}\!-\!k}\!\right)^{\!\frac{1}{2}}I_{1}\!\Big[\tilde k\cc,D(k,\tilde kr,q_l)\!-\!\frac{2}{k}\Big]~~~~~~~~~~
\frac{\Delta L}{L}\bigg|_{l}\!\approx  -\clM I_0(\tilde k r)H(k)\left(\!\frac{1-\tilde k\cc}{q_{l}\!-\!k}\!\right)^{\!\frac{1}{2}}I_{1}\!\left[\tilde k\cc,D(k,\tilde k r,q_l)\right],
\eeq
where $\tilde k\cc\equiv r\cc/r\per$ is essentially a function of $k$ only, as $\tilde k$. Hence, adding up the upper and lower parts of the integrals, we arrive at the desired fully analytic expressions of $\Delta E/E$ and $\Delta L/L$. 

It is worth mentioning that, in the cases mentioned in Section \ref{intro} that there is no pericentre, the solution of equation (\ref{eq}) is zero, implying that $r\cc$ is equal to the formal (see Sec.~\ref{intro}) lower limit $r\per$ of the integrals in equations (\ref{AE2}) and (\ref{AL2}), so the only contributions to $\Delta E/E$ and $\Delta L/L$ are from the upper parts.



\section{Evolved-to-Original Apocentric Radius and Mass Ratios}\label{App3}

The partial derivatives entering the definition of the function $\mu_{\rm DF}$ (eq.~[\ref{munew}]) are part of the Jacobian $J_{r\ff,\clM\ff}^{r,\clM}$ of the transformation: $r=r(v,r\ff,\clM\ff)$ and $\clM=\clM(v,r\ff,\clM\ff)$, i.e. the inverse of the Jacobian $J_{r,\clM}^{r\ff,\clM\ff}$ of the transformation $r\ff=r\ff(v,r,\clM)$ and $\clM\ff=\clM\ff(v,r,\clM)$ defined in Section \ref{multiple}. The latter Jacobian must be calculated recursively, like the functions $r\ii$ and $M\ii$ themselves, taking into account that, after the $i+1$ orbit, the Jacobian $J_{r,\clM}^{r_{i+1},M_{i+1}}$ is the matrix product of the Jacobian $J_{r,\clM}^{r\ii,M\ii}$ calculated in the previous step times the Jacobian $J_{r\ii,M\ii}^{r_{i+1},M_{i+1}}$ of the new elementary transformation $r_{i+1}=r_{i+1}(v,r\ii,M\ii)$ and $M_{i+1}=M_{i+1}(v,r\ii,M\ii)$, whose elements follow from the following partial derivatives (see eqs.~[\ref{Mi}]-[\ref{ri}])
\beq
\frac{\partial\, r_{i+1}}{\partial\, x\ii}=\frac{1}{2}\,\frac{\partial\{ r\ii \left[Q\ff(k\ii,r\ii,M\ii)+Q\ff(k\ii,r\ii,M_{i+1}\right]\}}{\partial x\ii}\qquad\qquad\qquad\qquad\qquad
\frac{\partial M_{i+1}}{\partial x\ii}=\frac{\partial
\left[M\ii \frac{f(c\ii Q_{i+1})}{f(c\ii)}\right]}{\partial x\ii}
\label{Mip1}
\eeq
where the function $x\ii$ stands for any of the two variables: $M\ii$ and $r\ii$. Of course, the partial derivative of each quantity with respect to $r\ii$ is the direct partial with respect to that variable $r\ii$ plus the partial derivative with respect to $k\ii$ times the partial of $k\ii=v^2/[v^2-2GM(r\ii)/r\ii]$ with respect to $r\ii$. On the other hand, as  $Q_{i+1}(k\ii,r\ii,M\ii)$ is the solution of the implicit equation (\ref{mtr2bis}), its derivatives can be obtained from derivation of that equation, leading to
\beq
\frac{\partial Q_{i+1}}{\partial x\ii}=
\frac{f(c\ii)Q_{i+1}^3\frac{\partial F(k\ii,r\ii,M\ii)}{\partial x\ii}}
{c\ii\frac{\der f(c\ii Q_{i+1})}{\der (c\ii Q_{i+1})}-3\frac{f\left(c\ii Q_{i+1}\right)}{Q_{i+1}}}
~~~~~~~~~~~~~~~~~~\qquad\qquad\qquad\qquad
F(k\ii,r\ii,M\ii)=\frac{f\left[c(r\ii) \tilde k\ii Q\ff(k\ii,r\ii,M\ii/2)\right]}{f[c(r\ii)]\tilde k\ii [Q\ff(k\ii,r\ii,M\ii/2)]^3}.
\eeq
(The derivatives $\der c\ii/\der x\ii$ are null because $c\ii$ is the initial concentration at the $i+1$ orbit, so it does not vary when the $x\ii$ values change.) 

To leading order in $\Delta E/E$ and $\Delta L/L$, the only non-null elements of the Jacobians $J_{\clM,r}^{M_{i+1},r_{i+1}}$ found at each step are those in the diagonal, which are precisely equal to the product of the corresponding elements of the two Jacobians that are multiplied. Consequently, the same is true for the final Jacobian $J_{\clM,r}^{\clM\ff,r\ff}$ whose diagonal elements take, to first order, the form (see eqs.~[\ref{lastr}]-[\ref{lastM}] and [\ref{radii}]-[\ref{finMp}])
\beqa
\frac{\partial r\ff}{\partial r}=\prod_0^\nu \frac{\partial r_{i+1}}{\partial r\ii}=\left(\prod_{i=0}^{\nu} \frac{r_{i+1}}{r\ii}\right)\!\left(1+ \sum_{i=0}^\nu \frac{\Delta r\ii}{r\ii}+\sum_{i=0}^\nu r\ii \frac{\partial}{\partial r\ii}\frac{\Delta r\ii}{r\ii}\right)\approx\frac{r\ff}{r}\left(1+\frac{\der\ln I_0}{\der \ln r}\right)\frac{\Delta r}{r}~~~~~~~~~~~~~~~~~~~~~~~~~~~~~~~~~~~~~~~~~~~~~\label{Drp}\\
\frac{\partial \clM\ff}{\partial \clM}=\prod_{i=0}^{\nu} 
\!\frac{\partial M_{i+1}}{\partial M\ii}=\!\Bigg(\prod_{i=0}^{\nu} \!\frac{M_{i+1}'}{M\ii}\Bigg)\Bigg(
1+2J(k,r)\sum_{i=0}^\nu\frac{\Delta r\ii}{r\ii} \Bigg)=\frac{\clM\tr}{\clM}\left[1+2J(k,r)\frac{\Delta r}{r}\right]\label{DMp}.~~~~~~~~~~~~~~~~~~~~~~~~~~~~~~~~~~~~~~~~~~~~~~~~
\eeqa
To derive equations (\ref{Drp}) and (\ref{DMp}) we have taken into account equations (\ref{Qf})-(\ref{rfr}) with $\Delta E/E$ and $\Delta L/L$ given in Appendix A, which leads to $\partial (\Delta r\ii/r\ii)/\partial M\ii=(\Delta r\ii/r\ii)/M\ii$ and 
\beq
\frac{\Delta r\ii}{r\ii}\!
=-\frac{k\ii H(k\ii)}{1\!-\!k\ii}M\ii\left[\!\frac {I_0(r\ii)}{S(k\ii,r\ii)}\left(\!\frac{1\!-\!\tilde k\ii}{1\!-\!k\ii}\!\right)^{\!\frac{1}{2}}\!\!\tilde I_{1}\!\left(\tilde k\ii\right)\!+\!\frac{I_0(\tilde k\ii r\ii)}{S(k\ii,\tilde k\ii r\ii)}\left(\!\frac{1\!-\!\tilde k_{{\rm ci}}}{q_{li}\!-\!k\ii}\!\right)^{\!\frac{1}{2}}\!\!\tilde I_{1}\!\left(\tilde k_{{\rm ci}}\right)\!\right],
\eeq
implying
\beq
r\ii\frac{\partial }{\partial r\ii}\!\frac{\Delta r\ii}{r\ii}\approx-\frac{k\ii H(k\ii)}{1\!-\!k\ii}M\ii\!\left[\frac{I_0(r\ii)}{S(k\ii,r\ii)}  \frac{\partial \ln I_0/S}{\partial \ln r\ii}\left(\!\frac{1\!-\!\tilde k\ii}{1\!-\!k\ii}\right)^{\!\frac{1}{2}}\!\tilde I_1(\tilde k\ii)
+\!\frac{I_0(\tilde k\ii r\ii)}{S(k\ii,\tilde k\ii r\ii)}\frac{\partial  \ln I_0/S}{\partial \ln (\tilde k\ii r\ii)}\left(\!\frac{1\!-\!\tilde k_{{\rm ci}}}{q_l\!-\!k\ii}\!\right)^{\!\frac{1}{2}}\!\tilde I_1(\tilde k_{\rm ci})\right]\approx \frac{\der \ln \tilde I_0}{\der \ln r\ii}\frac{\Delta r\ii}{r\ii},
\label{i}
\eeq 
where we have defined $\tilde I_{1}(\tilde k)\equiv 0.625(1-\tilde k)[1/1.875\!+\!3/5(1-\tilde k)]2/k$ and $\tilde I_0(r)\equiv j(r)I_0(r)$, with $j(r)\equiv f[c(r)]/\ln [1+c(r)]$ being a weak function of $r$, and we have neglected the logarithmic radial derivative of $(1-\tilde k)j(r)\der \ln \tilde I_0/\der \ln r$ in front of unity.

Equation (\ref{i}) leads to 
\beq
\frac{\partial }{\partial \ln r}\frac{\Delta r}{r}\approx \frac{\der \ln \tilde I_0}{\der \ln r}\frac{\Delta r}{r},\!
\label{rver}
\eeq 
whose solution is 
\beq
\frac{\Delta r}{r}\approx -Y(k)\tilde I_0(r)\clM\qquad~~~~~~~~~~~~~~~~~~~~~~~~~\qquad
Y(k)=\frac{k^2 H(k)}{2(1\!-\!k)}\left[\left(\!\frac{1\!-\!\tilde k}{1\!-\!k}\!\right)^{\!\frac{1}{2}}\!\!\tilde I_{1}\!\left(\tilde k\right)\!+\!(k-0.1)\left(\!\frac{1\!-\!\tilde k_{{\rm c}}}{q_l\!-\!k}\!\right)^{\!\frac{1}{2}}\!\!\tilde I_{1}\!\left(\tilde k_{{\rm c}}\right)\!\right].
\label{Y}
\eeq
Then, equation (\ref{DMp}) implies 
\beq
\frac{\partial \clM}{\partial \clM\ff}=\frac{\clM}{\clM\tr}\left[1-2J(k,r)\frac{\Delta r}{r}\right]\approx \frac{\partial \clM}{\partial \clM\tr}\left[1+2J(k,r)Y(k)\tilde I_0(r)\clM\right]\label{partial1}.
\eeq
On the other hand, differentiating $M(r\ff)$ with respect to $M(r)$ in the Taylor expansion 
\beq
M(r\ff)=M(r)\left[1+\frac{\der \ln M}{\der \ln r}\frac{\Delta r}{r}\right]\approx M(r)\left[1-\frac{\der \ln M}{\der \ln r}Y(k)\tilde I_0(r)\clM\right]
\eeq
and neglecting the double logarithmic derivative of $M(r)$, we obtain, to first order,
\beq
\frac{\partial M(r)}{\partial M(r\ff)}\approx 1+\frac{\der \ln (M\tilde I_0)}{\der \ln r\ff}Y(k)\tilde I_0(r\ff)\clM\ff.\label{second2}
\eeq

Equations (\ref{partial1}) and (\ref{second2}) give the two factors in the first term of the angular bracket in $\mu\DF$ (eq.~[\ref{munew}]), while the factors in the second term are null. Consequently, taking into account the expression of $\mu$ (eq.~[\ref{mu}]) and the relation (\ref{masses}), the average in equation (\ref{munew}) can be expressed, to first order, in the form
\beq
\mu\DF(r\ff,\clM\ff)\approx\mu(r\ff)+\omega(r\ff)\clM\ff  \qquad \qquad\qquad\qquad \qquad\qquad
\omega(r\ff)=\kappa(r\ff)\frac{\der \ln (M\tilde I_0)}{\der \ln r\ff} \tilde I_0(r\ff), 
\label{muDF}
\eeq
where $\mu(r\ff)\equiv \lav \clM\tr/\clM\rav(r\ff)$ (eq.~[\ref{mu}]) and $\kappa(r\ff)\equiv \lav Y \clM\tr/\clM\rav(r\ff)$. Note that the $v$-PDF is tiny near its upper bound (see Paper II), so the upper limit of the integral over $v$ defining the average in angular brackets, equal to first order to
\beq
v\maxi=\left[\frac{GM(r\ff)}{r\ff}\right]^{1/2}\left\{1+\frac{1}{2}\frac{\der \ln [M(r\ff)/r\ff]}{\der \ln r\ff} \frac{\Delta r}{r}(r\ff,\clM\ff)\right\},
\eeq
can be approximated by $[GM(r\ff)/r\ff]^{1/2}$, so angular brackets in equation (\ref{munew}) can be seen to denote velocity average for subhaloes at $r\ff$, as it does in equation (\ref{mu}) for subhaloes at $r$.

We remark that the change in the strength of tidal stripping and shock-heating due to DF (given by the term with $J(k,r)$ in eq. [\ref{masses}]) cancels to first order when deriving equation (\ref{muDF}), so the effect of DF on $\mu$ arises directly from the change in subhalo radii (through $\partial M(r)/\partial M(r\f)$) not from the change in the strength of tidal stripping and shock-heating (through $\partial \clM^{-1}/\partial (\clM\ff)^{-1}$). Certainly, for those values of $r$ or $k$ leading to spiral orbits without pericentre, subhaloes spiral inwards without tidal stripping, so one has $\clM\ff=\clM$, implying that the term with $J(k,r)$ is of order unity and cannot cancel. But, the fraction of subhaloes with any mass at very small $r$ is tiny and so is also the fraction with $k$ close to unity at any $r$. (Given the shape of the $v$-PDF, the fraction of subhaloes at each $r$ with  $k>0.9$ is much less than 10\%.) Therefore, we can safely ignore those cases and concentrate in those leading to equation (\ref{muDF}).

\end{document}